\documentclass[twocolumn]{revtex4}
\usepackage{}
\usepackage{epsfig,latexsym,amssymb}
\usepackage{latexsym}
\usepackage{amssymb}
\usepackage{amsfonts}
\usepackage{comment}
\usepackage{graphicx}
\usepackage{verbatim}

\newcommand{\be}{\begin{equation}}
\newcommand{\ee}{\end{equation}}
\newcommand{\bea}{\begin{eqnarray}}
\newcommand{\eea}{\end{eqnarray}}

\begin{document}

\title{Vector Dark Matter Detection using the Quantum Jump of Atoms}

\author{Qiaoli Yang$^{1,2,}$\footnote{qiaoli\_yang@hotmail.com} and Haoran Di$^{3}$}
\affiliation{$^{1}$Siyuan Laboratory, Department of Physics, Jinan University, Guangzhou 510632, China\\$^{2}$Department of Physics, Jinan University, Guangzhou 510632, China\\$^{3}$School of Physics, Huazhong University of Science and Technology, Wuhan 430074, China}

\begin{abstract}
The hidden sector U(1) vector bosons created from inflationary fluctuations can be a substantial fraction of dark matter if their mass is around $10^{-5}$eV. The creation mechanism makes the vector bosons' energy spectral density $\rho_{cdm}/\Delta E$ very high. Therefore, the dark electric dipole transition rate in atoms is boosted if the energy gap between atomic states equals the mass of the vector bosons. By using the Zeeman effect, the energy gap between the 2S state and the 2P state in hydrogen atoms or hydrogen like ions can be tuned. The $2S$ state can be populated with electrons due to its relatively long life, which is about $1/7$s. When the energy gap between the semi-ground $2S$ state and the 2P state matches the mass of the cosmic vector bosons, induced transitions occur and the 2P state subsequently decays into the 1S state. The $2P\to1S$ decay emitted Lyman-$\alpha$ photons can then be registered. The choices of target atoms depend on the experimental facilities and the mass ranges of the vector bosons. Because the mass of the vector boson is connected to the inflation scale, the proposed experiment may provide a probe to inflation.
\end{abstract}

\maketitle

\section{Introduction}
The existence of dark matter has been widely accepted due to the discovery of ample evidence such as the galactic rotational curves, the large scale structures, the gravitational lensings and the observations of the cosmic microwave background anisotropy etc. \cite{Battaglia:2005rj,Springel:2005nw,Smith:2006ym,Duffy:2008pz,Okabe:2009pf,Guo:2009fn,Postman:2011hg,Ade:2013zuv}. The properties of dark matter particles include that they are non-baryonic, weakly interacting and stable. There are many theories that can provide a proper dark matter candidate and a large part of these dark matter candidates can be categorized into two classes: 1, axions/axion like particles (ALPs) \cite{Ipser:1983mw,Preskill:1982cy,Abbott:1982af,Dine:1982ah,Khlopov:1999tm,Berezhiani:1990sy,Svrcek:2006yi,Arvanitaki:2009fg,Yang:2015qka,Arias:2012az} created by the misalignment mechanism and massive vector dark bosons \cite{Arvanitaki:2009hb,Nelson:2011sf,Graham:2015rva,Arias:2012az} created from the misalignment mechanism or inflationary fluctuations; and 2, weakly interacting massive particles (WIMPs) such as the TeV scale supersymmetric particles \cite{Jungman:1995df} created from the thermal production in hot plasma. The axions/ALPs and the vector dark matter are bosons with a typically smaller mass ($<$eV) and higher phase space density, which makes them behave more like waves or condensate. The WIMPs are much heavier ($>$GeV) and have a thermal distribution so they behave more like particles. Experiments searching for axions/ALPs, vector dark bosons, or WIMPs are currently proceeding or in planning in laboratories around the world \cite{Asztalos:2009yp,Wagner:2010mi,Hoskins:2011a,Irastorza:2011gs,Tam:2011kw,Arias:2012az,Horns:2012jf,Sikivie:2014lha,Jaeckel:2007ch,Hochberg:2015fth,Agnese:2014aze,Aprile:2012nq,Aprile:2017iyp,Essig:2012yx,Chaudhuri:2014dla,An:2014twa,Hochberg:2016ajh,Asztalos:2011bm}.

The hidden massive U(1) vector boson, dark photons, can be a substantial fraction of dark matter. The cosmic dark photon populations are generally non-thermally created by the misalignment mechanism and/or from inflationary fluctuations. The inflationary fluctuation creation of dark photons \cite{Ford:1986sy,Lyth:1996yj,Graham:2015rva} is appealing because it connects the dark matter mass with the Hubble scale of inflation. It is found that although the well known scalars and tensors power spectra created from the inflation fluctuations are scale invariant, the vector power spectrum peaks at intermediate wave length. Therefore, long-wavelength, isocurvature perturbations are suppressed so the production is consistent with the cosmic microwave background anisotropy observations.

The number density $N$ of sub eV dark photons is currently very high, of the order of $N=\rho_{cdm}/M\gtrsim 3*10^8/{\rm cm^3}$, where $\rho_{cdm}$ is the dark matter energy density. Therefore we can treat the cosmic dark photons as a classical field. The dark photon field is mostly composed by the dark electric filed $|\vec E'_0|\approx \sqrt{2 \rho_{cdm}}$, and in addition, the cosmic dark photons have a very high phase space density because their velocity dispersion is the order of $\delta v\sim v\sim 10^{-3}c$. Thus the electric dipole transition induced by the dark photons in an atom is enhanced. This makes the quantum transitions of atoms or ions a suitable method for detecting cosmic dark photons.

Many proposed and current experimental studies are looking for cosmic dark photons \cite{Jaeckel:2007ch,Arias:2012az,Horns:2012jf,Hochberg:2015fth,Wagner:2010mi,An:2014twa,Hochberg:2016ajh,Chaudhuri:2014dla,Asztalos:2011bm}. The proposed and current experiments include electromagnetic resonator experiments (such as the ADMX), LC oscillator experiments, Xenon10, and the newly proposed absorption of dark matter by a superconductor. Each experiment suits a different mass range. The proposed study presented here is suitable for $M\lesssim2*10^{-4}$eV with a higher sensitivity when the mass is smaller, please refer to Figure 3.

\section{Vector Dark Matter}
The hidden U(1) vector boson has a small mass and a very weak coupling to the standard model photon. Let us use $A'_{\mu}$ to denote the new vector field, the effective Lagrangian therefore can be written as:
\bea
{\cal L}&=&-{1 \over 4}(F^{\mu\nu}F_{\mu\nu}+F'^{\mu\nu}F'_{\mu\nu}+2\chi F'^{\mu\nu}F_{\mu\nu})\nonumber\\
&-&{M^2\over 2} A'_{\mu}A'^{\mu}-e\bar \psi \gamma^\mu\psi A_{\mu}+...  ~~,
\eea
where $F'_{\mu\nu}=\partial_{\mu} A'_{\nu}- \partial_{\nu} A'_{\mu}$, $\chi$ is the mixing parameter, $M$ is the mass of the hidden U(1) boson, and $\psi$ are fermions with ordinary electric charge in the standard model sector. The mixing term results in oscillations between the two U(1) bosons. We can redefine the field to mass eigenstates to get a massive vector boson and a massless vector boson without mixing up to $O(\chi^2)$:

\bea
A_{\mu} &\to&  A_{\mu}-\chi A'_{\mu} \nonumber\\
{\cal L}&=&-{1 \over 4}(F^{\mu\nu}F_{\mu\nu}+F'^{\mu\nu}F'_{\mu\nu})-{M^2\over 2} A'_{\mu}A'^{\mu}\nonumber\\
&-&e\bar \psi \gamma^\mu\psi A_{\mu}-\chi e\bar \psi \gamma^\mu\psi A'_{\mu}+...  ~~.
\eea
We see that the new massive vector boson, the dark photon, couples to the standard model charged fermions very weakly with an effective coupling constant $\chi e$. The value of the two parameters, the mass $M$ of the dark photon, and the coupling suppression factor $\chi$ are crucial to the phenomenologies of this model.

Cosmic dark photons can be created from inflationary fluctuations. Inflation during the early universe addresses many cosmological puzzles and is therefore a compelling model of the evolution of the universe \cite{Guth:1980zm,Linde:1981mu}. The inflationary fluctuation that produces dark photons is purely gravitational thus only requires the dark photons to couple to the standard model sector particles weakly to avoid over production in hot plasma. The large scale isocurvature perturbations of the dark photons are suppressed so the power spectrum is dominated by adiabatic perturbations, which is consistent with current observations. The abundance of dark matter in this scenario is determined by the Hubble scale of inflation and the mass of dark photons:
\be
\Omega_{A'}/\Omega_{cdm}=[M/(6*10^{-6}{\rm eV})]^{1/2}\times [H_I/10^{14}{\rm GeV}]^2~~,
\ee
where $H_I$ is the Hubble scale of inflation.

The cosmic dark photons are currently free streaming. Using the Lorentz gauge condition
\be
\partial_{\mu}A'^{\mu}=0~~,
\label{constrain}
\ee
then the field obeys the wave equation: $(\partial_{\mu}\partial^{\mu}+M^2)A'_{\mu}=0$. As the cold dark matter particles are non-relativistic, in the momentum space we have:
\be
A'_{\mu}(\vec v,t)\approx A'_{\mu}e^{i(-Mt-{M\over 2}v^2t+M\vec v\cdot \vec x)}~~,
\label{EM}
\ee
up to the second order of velocity $v$. From Eq.(\ref{constrain}) and Eq.(\ref{EM}) we find that the time component of the vector field is suppressed by velocity $v$ and is therefore small. For our subsequent discussions it is convenient to use the dark electric field $\vec E'$ and dark magnetic field $\vec B'$ instead of the vector field $A'_{\mu}$. Because the spacial part of the vector field is much larger than the time part, we have $\vec E'=-\partial \vec A'/\partial  t\approx -iM\vec A'$ and $\vec B'=\nabla \times \vec A'\approx 0$. The energy distribution is:
\bea
I_{A'}={\rho_{cdm}\over \Delta E}\approx {0.3{\rm GeV/cm^3}\over (1/2)M\Delta v^2}={6*10^5\over Mc^2}{\rm GeV/cm^3},
\label{intensity}
\eea
where $\Delta v\sim 10^{-3}c$ is the typical estimate of the cold dark matter velocity distribution. As $\Delta v\approx 2\sqrt{T/M}$, where $T$ is the effective temperature of the dark matter, the energy distribution is higher when the dark matter is colder. Literature \cite{Armendariz-Picon:2013jej} finds $T_{today}/M\sim 10^{-14}$ which corresponds to a $\Delta v\sim 10^{-7}c$. This result will boost the number of events or the signal of our experiment order of $10^8$ comparing to the $\Delta v\sim 10^{-3}c$ case (see Eq.(\ref{NET})). In the following discussions, we still use the more conservative estimation $\Delta v\sim 10^{-3}c$.

\section{Design of the Experiment}
The hidden photon couples to fermions via:
\be
{\cal L}_{\bar \psi\psi A'}=-\chi e\bar \psi \gamma^\mu\psi A'_{\mu}~~,
\label{interaction}
\ee
where $\psi$ is the electron field and $\chi$ is generally suppressed by loops in a more fundamental theory. The dark photons created from inflationary fluctuations have a mass of $10^{-5}$eV if they are a major part of the dark matter. However, the creation mechanism itself puts little constraint on the coupling $\chi$.

The Compton wavelength of the dark photon is $\lambda =2\pi (M)^{-1}$. If we use the standard assumption that $M\sim 10^{-5}$eV, the wave length is much larger than the Bohr radius $a_0\approx 5*10^{-11}$m of atoms. Therefore the dark electric field can be treated as a homogeneous field in atoms:
\be
|\vec E|=\sqrt{2\rho_{cdm}}cos(Mt)~~.
\ee
In the non-relativistic limit, Eq.(\ref{interaction}) leads to the Hamiltonian:
\be
H=-\chi e (\vec E' \cdot \vec x)-[\chi e/(4M)]\vec \sigma\cdot \vec B'+...~~,
\ee
where $\sigma$ is the Pauli matrices. We see that the first term is similar to the coupling of the electric dipole interaction and the second term plays the role of the magnetic momentum interaction. The second term is negligible when the dark magnetic field is small. The dark dipole coupling of atoms cause $\Delta l=\pm 1$, $\Delta m=0,~\pm 1$ transitions if the energy gap between two states matches the energy of the dark photons, where $l$ is the orbital angular momentum and $m$ is the third component of angular momentum. The energy gap between two states can be adjusted by using the Zeeman effect with an external magnetic field $\vec B$. The general Hamiltonian of the Zeeman effect is $H=-\vec \mu\cdot \vec B$, where $\vec \mu$ is the magnetic moment of the electron. The mass range that can be scanned is limited by the available magnetic field strength. Given today's technology, $B\sim 18$T \cite{Petrakou:2017epq}, we have $M\sim 240$GHz.
\begin{figure}
\begin{center}
\includegraphics[width=0.5\textwidth]{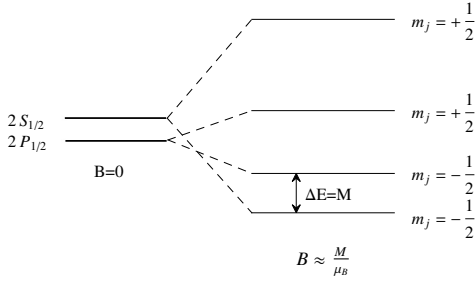}
\caption{The Zeeman effect on the 2S state and the 2P state. The energy gap between two states can be tuned using an external magnetic field. If the energy gap between two states matches the dark photon's mass, resonance transitions will occur.}
\end{center}
\end{figure}

The transition rate $R$ of atoms or ions from an initial state $|~i>$ to an excited state $|~f>$ is
\be
R={2\pi \chi ^2 e^2}{<|\vec E'_0|^2>\over max(\Delta \omega_{A'},\Delta \omega_{if},\Delta\omega)}|~\vec r_{i,~f}~ |^2~~,
\ee
where $|\vec r_{i,~f}|$ is the quantum matrix element between states $|~i>$ and $|~f>$, $\Delta \omega_{A'}={1\over 2}M\Delta v^2$ is the bandwidth of cosmic dark photons, $\Delta \omega_{if}={1/ \tau}$ is the bandwidth of the excited state, $\Delta \omega={1/ \Delta t}$ is the bandwidth of the useful integration time in a particular frequency range and $<|\vec E'_0|^2>={|\vec E'_0|^2\over 3}$ means a spacial average of the field. The resonance condition is $M=E_f-E_i$ and $E_i,~E_f$ are the energies of the initial state and the final state, respectively. Because for the experiment, $\Delta \omega_{A'}\gg\Delta \omega\gg\Delta \omega_{if}$, we have:
\be
R={4\pi \chi ^2 e^2\over 3}I_{A'}|~\vec r_{i,~f}~ |^2~~,
\ee
where the $I_{A'}$ is defined by Eq.(\ref{intensity}) which is the local dark photon energy spectrum distribution.
The exact value of the matrix element of dipole transition $|~\vec r_{i,~f}~|^2$ depends on the particular target material but we can estimate the order of magnitude in this preliminary assessment, which is considered a $2S\to 2P$ transition:
\be
|~\vec r_{i,~f}|^2\sim a_0^2~~.
\ee
The number of excited atoms or the number of events will be:
\bea
RNt={4\pi\over 3}  \chi^2e^2I_{A'}a_0^2Nt=1.93*10^8\chi^2N{(t/{\rm second})\over (M/{\rm eV})}
\label{NET}
\eea
where $N$ is the number of populated $2S$ states and $t$ is the integration time. For a case that $N\sim 10^{-6}$mole, $\chi\sim 10^{-15}$ and $M\sim 10^{-5}$eV, we have the number of events is 11.6 per second. These excited $2P$ atoms will decay rapidly into $1S$ atoms, and the emitted Lyman-$\alpha$ photons can be registered as the number of events.

Because the electric dipole transition of $2S\to 1S$ is forbidden, the $2S$ state is semistable with a lifetime of about 1/7s, which is much larger than the lifetime, $2*10^{-11}$s, of the $2P$ states. The $2S$ semistable states can be populated with electrons \cite{twophotonexc}. Let us assume $10^{-5}{\rm mole/ sec}$ of the $2S$ state are excited, which takes order of 1W power, then the populated $2S$ states are about $10^{-6}$ mole at any given time.

The set-up of the proposed experiment can be very similar to the experiments measuring the Lamb shift or the 1S-2S transition frequencies of atoms \cite{Niering:2000ab}. A major difference between the experiments is that for the existing experiments, microwaves are used to stimulate transitions between the $2S$ and the $2P$ state while in the proposed experiment the cosmic dark photons stimulate the $2S\to 2P$ transitions. Please refer to Figure 2 for a conceptual set-up. The cooled hydrogen atomic beam enters an interaction region which is a laser enhancement cavity with a doppler free standing laser wave near 243 nm. In the interaction region the atoms are excited by two-photon spectroscopy from the $1S$ ground state to the excited $2S$ metastable state. The atoms then enter the detection region where an external magnetic field adjusts the energy gap between the $2S$ and the $2P$ state. If the energy gap matches the dark photons' mass, the atoms will be stimulated to the $2P$ state and then decay into the $1S$ state with emitted Lyman-$\alpha$ photons which can be detected by a photomultiplier. The dark count rate of cooled photomultipliers operating at the optimal frequency can be very low which is order of a few hertz, thus we expect that a photon detection efficiency order of 0.6 can be achieved assuming that the total solid angle is covered. If, however, only a $\Omega$ solid angle is covered by the photomultipliers, the efficiency will be reduced by a factor of $\Omega/(4\pi)$.

The major noise of the proposed experiment comes from the thermal photon induced $2S\to 2P$ transitions. When the signal to noise ratio is bigger then one, we have:
\be
I_{t}(M)={\omega_A^3\over \pi^2c^3}{1\over {\rm exp}[{\hbar\omega_A\over k_BT}]-1}<\chi^2I_{A'}~~,
\label{themal}
\ee
where $I_{t}$ denotes the thermal photon energy distribution and $\omega_A=M$ is the frequency of the thermal photons. Eq.(\ref{themal}) leads to:
\be
T_{optimal}\leq {1.16*10^4({M\over{\rm eV}})\over {\rm ln}[{1\over 45.3\chi^2}({M\over{\rm eV}})^4+1]}{\rm K}~~,
\ee
which is the optimal working temperature of a dark-photon detection experiment. A possible method to produce the required low-temperature atoms can be the laser cooling technology such as described in \cite{lasercooling} and currently $10^{10}$ to $ 10^{12}$ cooled atoms can be produced per second by small-compact devices using only 20mW power \cite{lasercooling2,lasercooling3,lasercooling4}. An atomic funnel similar to \cite{atomicfunnel,atomicfunnel2} may be used to produce the injection atomic beam. The available number of cold atoms, $N$, could be a potential limitation for achieving a high sensitivity but fortunately the sensitively $\chi\propto 1/ \sqrt N$ as we will show in section IV so a moderate decreasing of the cold atom numbers could be affordable. To achieve a sensitivity $\chi\geq 10^{-17}$ with $M\sim 10^{-5}$eV, the optimal temperature is $4.08$mK and when $M\sim 10^{-4}$eV, the optimal temperature is $30.82$mK. A several mK temperature can be achieved for Hydrogen atoms according to \cite{Setija:1993zz}. If the achievable temperature is higher than the optimal temperature, a detection can be achieved in an expense of a longer integration time. As the thermal photon induced transition rate is $4\pi e^2I_{t}a_0^2/3$, a 95\% confidence detection requires $\rm {signal/ \sqrt{noise}}>3$, where ${\rm signal}=RNt$ and ${\rm noise}=R_{t}Nt$ respectively, so a detection requires $R/R_{t}^{1/2}*(Nt)^{1/2}>3$ if $T>T_{optimal}$.

\begin{figure}
\begin{center}
\includegraphics[width=0.5\textwidth]{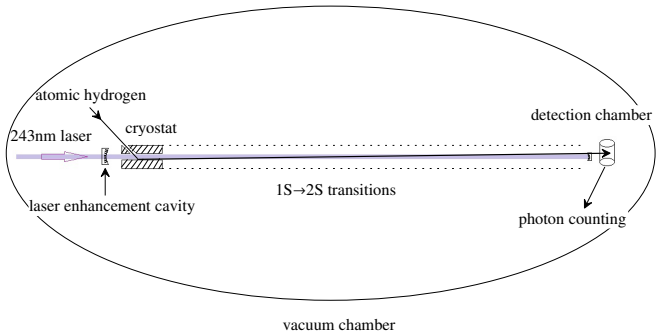}
\caption{A conceptual set-up of the proposed experiment.}
\end{center}
\end{figure}

\section{Sensitivity}
The sensitivity of the experiment depends on the integration time, thermal noises and number of cooled atoms. Let us assume a frequency bandwidth $\Delta B=M/(2\pi)$ is covered per working year for each experiment cycle. Then the magnetic field to induce the Zeeman effect is tuned so that the energy gap between two relative atomic states is shifting as:
\be
{\Delta B \over t_{cy}}={M/(2\pi)\over 1~{\rm year}}=77~{\rm Hz \over sec}({M\over{\rm 10^{-5} eV}})~~.
\ee
Because the band width of cosmic dark photons is $\Delta \omega_{A'}=(M\Delta v^2)/2$, during a cycle the event integration time is $(\Delta \omega_{A'}/\Delta B)*t_{cy}=3.14*10^{-6}t_{cy}$.

During each cycle of the experiment, counted events can be checked by temporarily staying the frequency tune to see if additional events are registered. Let us use $\eta$ to denote the efficiency of the photon detector in counting an actual event. When the detector is working at the optimal temperature, to have a 95\% confidence detection, the registered number of events satisfy $NRt>3/\eta$. The sensitivity of the coupling $\chi$ is then:
\bea
\chi&>&{3\over 2a_0e}({1\over\pi I_{A'}N\eta}*{\Delta B\over t_{cy}\Delta\omega_{A'}})^{1/2}\nonumber\\
&=&1.25*10^{-5}({M\over {\rm eV}})^{1/2}\left({t_{cy}\over {\rm 1year}}*N\eta\right)^{-1/2}~~.
\eea
For a preliminary set up with $10^{-6}$ mole $2S_{1/2}$ atoms, one year cycle time, and a detection efficiency $\eta\sim 0.6$, the sensitivity is $\chi\sim 6*10^{-17}$ for $M\sim 10^{-5}$eV, please refer to Figure 3.
\begin{figure}
\begin{center}
\includegraphics[width=0.5\textwidth]{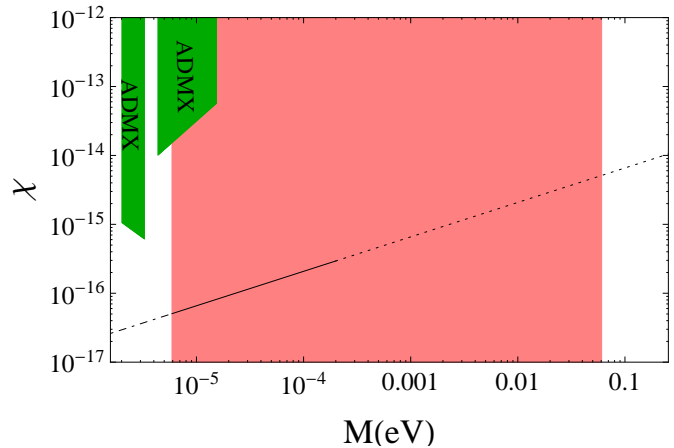}
\caption{Expected sensitivity of the experiment. The vertical pink band represents the possible mass range of the dark photon created by inflation fluctuations. The inflation production mechanism is completely gravitational therefore does not have a theoretical constraint on the coupling of dark photons. The area above the dark line is the sensitivity region for the preliminary set up of the experiment. The green regions are excluded by current results from the ADMX dark matter searches \cite{Asztalos:2011bm}. The left side dot-dashed line means that the dark photons may be created from other mechanism instead of inflation fluctuations but we still assume that they are a substantial part of dark matter. The right side dotted line means that the experiment can only partially cover the mass range due to the available magnetic field strength limitation $\sim 18$T.}
\end{center}
\end{figure}
\section{Conclusions}
The hidden sector is a natural extension of the Standard Model of particle physics. Most models with a hidden sector include gauge groups that are independent from the known $U(1)\times SU_L(2)\times SU_C(3)$ standard model gauge groups. Therefore, hypothetical particles in the hidden sector interact very weakly with the standard model particles.

If the new U(1) massive dark photons exist, they can be naturally produced by inflation. The production does not ruin the CMB power spectrum. In addition, the production mechanism does not need a specified model because it is completely gravitational. Thus, the abundance of the production only depends on the Hubble scale of inflation and the mass of the dark photon. Given the high energy scale of inflation $\gtrsim 10^{14}$GeV, and the rich ultraviolet structures of such a high energy scale, the uncertainty of the coupling is very high.

Experimental detections of these particles can serve as a probe to inflation. There are two practical problems in such an experiment: the first is that the coupling is very weak and the second is that the range of mass is very wide. Therefore, to cover the parameter space as much as possible, multi-type experiments may be needed. In this paper, we propose the use of atomic transitions to detect the vector boson dark matter. The high energy spectral density of the vector boson dark matter will boost the transition rate of atoms if the energy gap between atomic states, which can be adjusted by the Zeeman effect, matches the mass of the dark photon. The excited states of the atoms then can be counted by registering the emitted Lyman-$\alpha$ photons. The reachable mass range of the experiment depends on the choice of target material and the available magnetic field for the Zeeman effect.
\bigskip

{\itshape Acknowledgments:} Q.Y. would like to thank Pierre Sikivie, Weitian Deng, Yungui Gong, Fen Zuo, Yunqi Liu, Bo Feng and Jianwei Cui for their helpful discussions. This work is partially supported by the Natural Science Foundation of China under grant Number 11305066.

\end{document}